# Kinetic proofreading of gene activation by chromatin remodeling

**R. Blossey***

**Interdisciplinary Research Institute CNRS USR 3078, Lille University of Science and Technology, c/o IEMN UMR 8520, Cité Scientifique BP 60069, F-59652 Villeneuve d'Ascq, France**

**H. Schiessel**

**Instituut Lorentz, Universiteit Leiden, P.O. Box 9506, 2300 RA Leiden, The Netherlands**

**\* corresponding author, ralf.blossey@iri.univ-lille1.fr**

## Abstract

Gene activation in eukaryotes involves the concerted action of histone tail modifiers, chromatin remodelers and transcription factors, whose precise coordination is currently unknown. We demonstrate that the experimentally observed interactions of the molecules are in accord with a kinetic proofreading scheme. Our finding could provide a basis for the development of quantitative models for gene regulation in eukaryotes based on the combinatorical interactions of chromatin modifiers.

## Introduction

How genes are activated in eukaryotic genomes is not yet fully understood, but it is becoming increasingly clear that the modulation of chromatin structure plays a crucial role. There is mounting experimental evidence that a strong correlation exists between histone tail modifiers, nucleosome remodelers and transcription factors (Strahl, B. D. et al., 2000, Horn P. J. et al., 2002, Schreiber S. L. et al, 2002, Cosgrove M.S. et al, 2204, Lue, N. F., 2005, Giresi P. G. et al, 2006). A quantitative (i.e., mathematical) description of the how these factors combine to address and activate genes is not presently available; but many qualitative models have been and are being deduced from experimental observations, as to be found in the above cited papers. Clearly, the three kinds of molecules known to intervene in chromatin regulation can be combined in various ways, and a specific experimental context may hide an underlying scheme in details specific to this particular biological model.

Our intention in this paper is twofold: i) to present a mechanism which may play a relevant role in the orchestration of chromatin regulators, and ii), to postulate a model which is sufficiently quantitative to validate or rule out some of the experimentally proposed scenarios. Motivated by experimental findings but abstracting from them, we have observed that the combined action of histone tail modifiers, chromatin remodelers and transcription factors allows for a kinetic proofreading scheme much akin to the originally proposed scheme (Hopfield, J. J. 1974), as we describe below. In the context of chromatin remodeling, such a scheme would allow to discriminate 'right' from 'wrong' genes to be activated for transcription at a pre-transcriptional level.

**Analysis**

Starting point for our model is the sequence of actions

$$\text{histone tail modifiers} \rightarrow \text{nucleosome remodellers} \rightarrow \text{transcription factors}$$

as displayed schematically in Figure 1. Throughout this paper we focus on acetylated nucleosomes, but this is no restriction; in reality much more complex histone tail modifications may be involved, and they can likewise be integrated into our model.

The sequence of activation steps we selected represents one example of several schemes discussed in the literature. It corresponds essentially to the scenario of Figure 1 C in (Eberharter A. et al., 2005), with the difference that we have placed the transcription factor at the end of the gene activation process; we comment on this difference at the end of the paper. For simplicity we only consider a two-state case, a fully aceytlated and a fully deacetylated nucleosome.

The overall reaction scheme we propose is presented in Figure 2.

Each kinetic proofreading scheme is based on equilibrium and non-equilibrium reactions (Eberharter A. et al, 2005). For chromatin, the equilibrium step is provided by the association of an acetylated nucleosome of concentration [A] with a remodeler [R] via the reaction [A] + [R] $\leftrightarrow$ [AR] with the on rate $r_+$ and the off-rate $r_-^A$. Second, we assume that the kinetics of the non-equilibrium remodeling reaction is governed by the scheme [AR] $\rightarrow$ [AR*] $\rightarrow$ [A] + [R] with reaction rates a and $d^A$, respectively. The *-symbol indicates the activated, let say more mobile nucleosome. Here, a and $d^A$ are the rates of nucleosome activation and of dissociation of the activated complex, respectively. The main input for postulating these two reactions is the knowledge that chromatin remodelers in general have different domains interacting with tail modifications and allowing for the processing of ATP. ATP is involved in our scheme both in fueling the nucleosomal mobility and in the proofreading discrimination mechanism.

In a quasi-steady state assumption, the ratio of concentrations of activated and non-activated nucleosomes is given by [AR*]/[AR] = $a/d^A$. In the final step, a transcription factor T can associate itself with the activated nucleosome with the reaction [AR*] + [T] $\leftrightarrow$ [AT] with on-rate $t_+$ and off-rate $t_-^A$.

For the deactylated nucleosome of concentration [D] we postulate a corresponding set of reactions (see Figure 2) with the same on-rates $r_+$ and $t_+$ for binding of the remodeler and the transcription factor as for the acetylated complex. Also, the rate a for the remodeling reaction is the same. However, we assume that the off-rates are different and indicate those of the deacetylated complex by an upper index D, $r_-^D$, $d^D$ and $t_-^D$, as compared to the rates $r_-^A$, $d^A$ and $t_-^A$ of the acetylated nucleosome.

As a measure of the efficiency of the proofreading scheme we consider the rate of formation of the final product, the accessibility of the readout gene G. For the acetylated case we assume this to be [AT]g and for the deacetylated case [DT]g, cf. Figure 2. We thus have to evaluate the ratio

$$R = [AT]/[DT] = (t_-^D \, d^D \, r_-^D \, k^D)/(t_-^A \, d^A \, r_-^A \, k^A) \tag{1}$$

Here, in the last step, the kinetics for the formation of the acetylated complex has been added, cf. Figure 2. Equation (1) is our central result. It describes the discrimination power of nucleosome activation associated with two different histone-modified states.

The discrimination of the acetylated state A and the deacetylated state D would be relatively small if it would rely only on the simple difference $\Delta U_T$ in binding energies of the transcription factor. Suppose that this affects its off-rates as

$$t_{-}^{D} = t_{-}^{A} \exp (\Delta U_T/k_B T) \qquad\qquad (2)$$

then the ratio R = [AT]/[DT] acquires a factor $\exp(\Delta U_T/k_B T)$. This factor alone would usually be not sufficient for discrimination.

The main effect of the proposed scheme lies in the recruitment of the remodeler and the nonequilibrium reaction, as is common for the kinetic proofreading scheme. We have

$$r_{-}^{D} = r_{-}^{A} \exp (\Delta U_R/k_B T) \qquad\qquad (3)$$

and

$$d^{D} = d^{A} \exp (\Delta U_R/k_B T) \qquad\qquad (4)$$

where $\Delta U_R$ denotes the energetic difference between the acetylated and deacetylated nucleosome-remodeler complexes. This means that the remodeling step leads to an extra factor $\exp(2 \Delta U_R/k_B T)$ for the ratio R = [AT]/[DT], eq.(1). Combining these expressions we find

$$R = [AT]/[DT] = \exp((2\Delta U_R + \Delta U_T)/ k_B T) (k_D/k_A)$$

Suppose one has only moderate effects of the acetylation on the binding energies, e.g. $\Delta U_T = \Delta U_R = 2k_B T$. Then we find nevertheless that the expression level of the acetylated region as compared to that of the deacetylated is dramatically enhanced by a factor $\exp((2 U_R + \Delta U_T)/k_B T \approx 400$. Without the remodeling step, however, one would only have $\exp(\Delta U_T/k_B T) \approx 7$.

**Application to the RSC remodeling system**

We finally confront our model with recently obtained experimental results on the RSC chromatin remodeling complex in yeast (*S. cerevisiae*). This complex is an especially well-characterized remodeling system to which a variety of techniques from molecular biology, biochemistry, single-molecule analysis, and electron microscopy have been applied, particularly by the Kornberg and Cairns groups (Asturias, F. J. et al., 2002, Saha A., et al, 2002, Kasten M. et al, 2004, Chai B. et al, 2005, Leschziner A. E. et al, 2007). Since structural knowledge of the remodeler-nucleosome interaction is becoming available recently (Asturias F. J. et al, 2002, Leschziner A. E. et al, 2007) we expect that an integrated model combining the effects of enzymatic modifications and the mechanistic aspects of chromatin remodeling may become available in the near future for this exemplary case.

A recent detailed biochemical analysis of the role of histone tail aceylation in chromatin remodeling in this system studied the interaction between the Rsc4 tandem bromodomain and the histone H3 tail acetylated at lysine 14 (Kasten M. et al, 2004). The authors studied in particular the role of bromodomain modifications and H3 Lys 14 substitutions. The authors considered their findings paradoxical: the presence of functional bromodomains turned out to be essential, while Lys14 – acetylation was not, since substituted H3 Lys14 were viable. Two interpretations were offered for this result: (i) Rsc4 bromodomain mutants may have a reduced binding interaction with both Lys 14 and flanking residues, so that acetylation at H3 Lys 14 may be needed for a sufficient binding level; (ii) Rsc4 bromodomains may bind two different targets on the nucleosome, one of which being H3 Lys 14Ac while the other is unknown. Both targets then together should provide sufficient binding energy.

We believe that these findings are much simpler to explain in the context of our model. The explanations of the authors of the study rely on the equilibrium binding energy of the remodeler only. It is clear from our model derived in the previous section that the main contribution in discrimination is due to the presence of the remodeling step, while changes in mere equilibrium binding matter considerably less. Therefore, the experimental observation can easily be interpreted by noting that the changes in binding energy due to the mutations brings the discrimination ratio minimally down for the mutant case, while the remodeling step itself still remains intact, and hence the cells viable.

The fact that also in the case of H3 Lys 14 substitutions cells are still viable (for the intact remodeler) raises the interesting question whether discrimination between acetylated and deacetylated nucleosomes is really crucial. But if the bromodomains in the absence of Lys 14 bind to other acetylated lysines as alternative targets, then a rather high discrimination ratio can still be achieved, again due to the presence of the remodeling step.

A fully quantitative application (and thereby validation) of our model necessitates a more complete knowledge of both the intervening molecular partners and their reaction rates, however, even in the absence of such knowledge it already puts constraints on the comparison of alternative scenarios. As such, it may also guide experimentation towards a verification of particular remodeling schemes.

**Discussion**

The kinetic proofreading scenario for chromatin remodeling presented here shows that chromatin remodeling based on histone tail modifiers, chromatin remodelers and transcription factors allows to discriminate 'right' from 'wrong' genes to be activated. While in our example we have distinguished between an acetylated vs. a deacetylated state, other covalent modifications and their combinations (the 'histone code', (Strhl B. D. et al, 2000) can clearly be included since their presence affects the off-rates $D$, $r_-^D$, $d^D$ and $t^D$. For each existing gene/promoter there is a combination of remodeling events which promotes gene access with the highest possible discrimination. We further stress that how these factors are scheduled is less essential: note that the final result eq.(1) does not change if the transcription factor is placed at the start of the initiation process. Finally we note that eukaryotic transcriptional activation usually involves a large number of transcription factors and gene regulatory proteins. Any component that is sensitive to histone modifications constitutes an additional factor in the discrimination power of eq.(1). The coupling of a remodeling step to any of those components, as exemplified above in our paper, leads to a dramatically increased sensitivity of that component to histone modifications.

**Figure Legends**

Figure 1: The sequence of regulatory events supposed in our model (schematic). Top: nucleosome with one histone tail carrying amino acid modifications. Middle: recruitment of a remodeling complex (R) with a histone modifier recognition unit (dark green) and the ATP-active domain (red). The DNA is partially loosened from the histone octamer. Bottom: recruitment of a transcription factor (T).

Figure 2: Kinetic proofreading scheme for the regulation of transcription initiation in chromatin. See text for details.

Figure 1

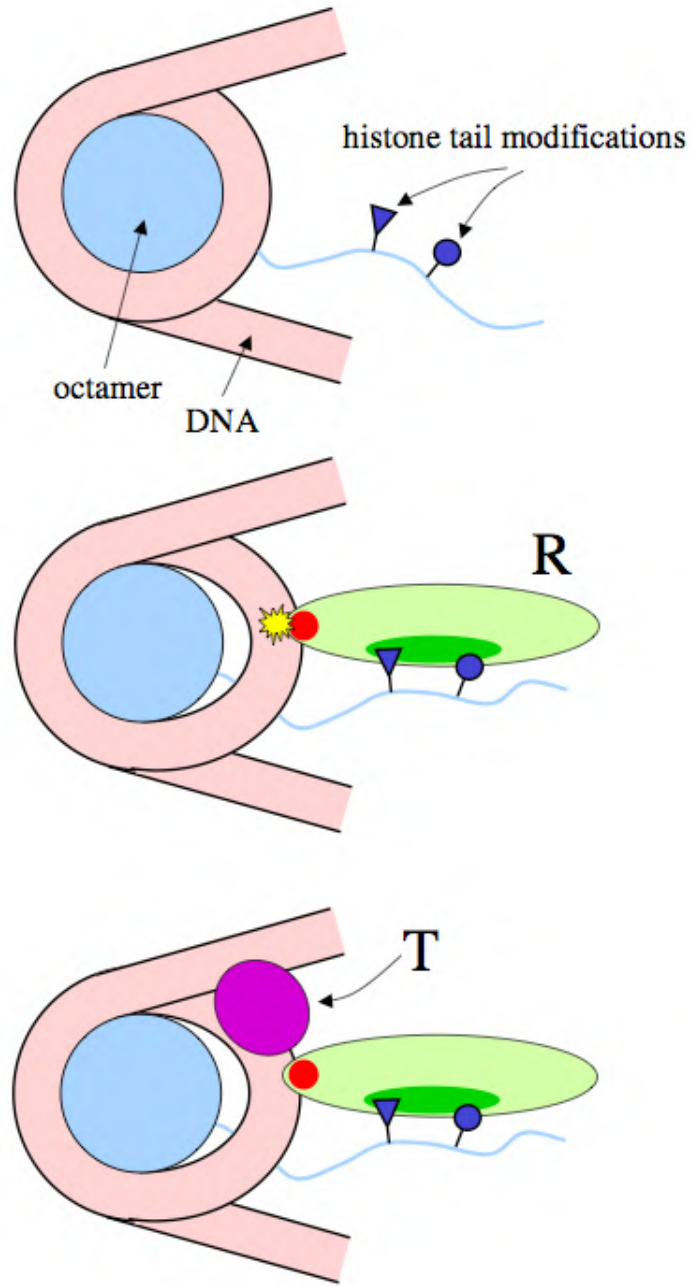

Figure 2

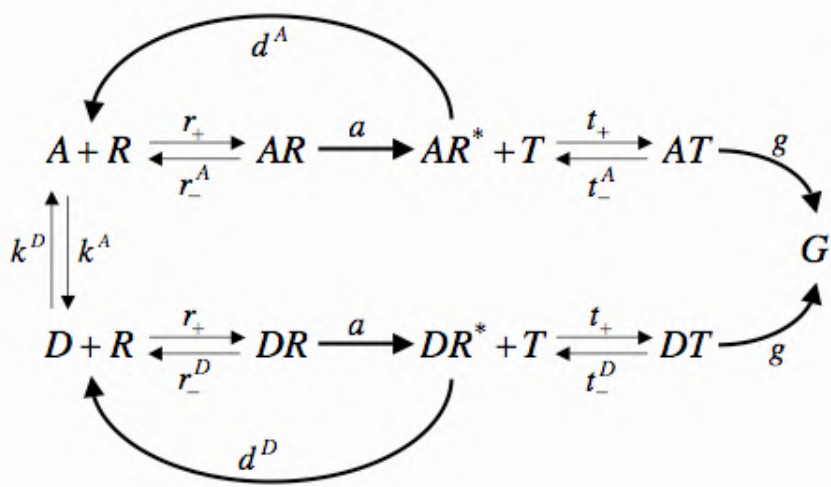